\RequirePackage{lineno}
\documentclass[preprint,amsmath,amssymb,superscriptaddress]{revtex4-2}
\usepackage{graphicx}
\usepackage{color}
\usepackage{newfloat}
\begin{document}

\title{Material design with the van der Waals stacking of bismuth-halide chains realizing a higher-order topological insulator}

\author{Ryo~Noguchi}
\affiliation{Institute for Solid State Physics, University of Tokyo, Kashiwa, Chiba 277-8581, Japan}

\author{Masaru~Kobayashi}
\affiliation{Materials and Structures Laboratory, Tokyo Institute of Technology, Yokohama, Kanagawa 226-8503, Japan}

\author{Zhanzhi~Jiang}
\affiliation{Department of Physics, The University of Texas at Austin, Austin, Texas 78712, United States}

\author{Kenta~Kuroda}
\affiliation{Institute for Solid State Physics, University of Tokyo, Kashiwa, Chiba 277-8581, Japan}

\author{Takanari~Takahashi}
\affiliation{Materials and Structures Laboratory, Tokyo Institute of Technology, Yokohama, Kanagawa 226-8503, Japan}

\author{Zifan~Xu}
\affiliation{Department of Physics, The University of Texas at Austin, Austin, Texas 78712, United States}

\author{Daehun~Lee}
\affiliation{Department of Physics, The University of Texas at Austin, Austin, Texas 78712, United States}

\author{Motoaki~Hirayama}
\affiliation{RIKEN Center for Emergent Matter Science (CEMS), Wako, Saitama 351-0198, Japan}

\author{Masayuki~Ochi}
\affiliation{Department of Physics, Osaka University, Toyonaka, Osaka 560-0043, Japan}

\author{Tetsuroh~Shirasawa}
\affiliation{National Metrology Institute of Japan, National Institute of Advanced Industrial Science and Technology, Tsukuba, Ibaraki 305-8565, Japan}

\author{Peng~Zhang}
\affiliation{Institute for Solid State Physics, University of Tokyo, Kashiwa, Chiba 277-8581, Japan}

\author{Chun~Lin}
\affiliation{Institute for Solid State Physics, University of Tokyo, Kashiwa, Chiba 277-8581, Japan}

\author{C\'edric~Bareille}
\affiliation{Institute for Solid State Physics, University of Tokyo, Kashiwa, Chiba 277-8581, Japan}

\author{Shunsuke~Sakuragi}
\affiliation{Institute for Solid State Physics, University of Tokyo, Kashiwa, Chiba 277-8581, Japan}

\author{Hiroaki~Tanaka}
\affiliation{Institute for Solid State Physics, University of Tokyo, Kashiwa, Chiba 277-8581, Japan}

\author{So~Kunisada}
\affiliation{Institute for Solid State Physics, University of Tokyo, Kashiwa, Chiba 277-8581, Japan}

\author{Kifu~Kurokawa}
\affiliation{Institute for Solid State Physics, University of Tokyo, Kashiwa, Chiba 277-8581, Japan}

\author{Koichiro~Yaji}
\affiliation{Institute for Solid State Physics, University of Tokyo, Kashiwa, Chiba 277-8581, Japan}

\author{Ayumi~Harasawa}
\affiliation{Institute for Solid State Physics, University of Tokyo, Kashiwa, Chiba 277-8581, Japan}

\author{Viktor~Kandyba}
\affiliation{Elettra - Sincrotrone Trieste, S.S.14, 163.5 km, Basovizza, Trieste, Italy}

\author{Alessio~Giampietri}
\affiliation{Elettra - Sincrotrone Trieste, S.S.14, 163.5 km, Basovizza, Trieste, Italy}

\author{Alexei~Barinov}
\affiliation{Elettra - Sincrotrone Trieste, S.S.14, 163.5 km, Basovizza, Trieste, Italy}

\author{Timur~K.~Kim}
\affiliation{Diamond Light Source, Harwell Campus, Didcot, OX11 0DE, United Kingdom}

\author{Cephise~Cacho}
\affiliation{Diamond Light Source, Harwell Campus, Didcot, OX11 0DE, United Kingdom}

\author{Makoto~Hashimoto}
\affiliation{Stanford Synchrotron Radiation Light source, SLAC National Accelerator Laboratory, 2575 Sand Hill Road, Menlo Park, California 94025, USA}

\author{Donghui~Lu}
\affiliation{Stanford Synchrotron Radiation Light source, SLAC National Accelerator Laboratory, 2575 Sand Hill Road, Menlo Park, California 94025, USA}

\author{Shik~Shin}
\affiliation{Institute for Solid State Physics, University of Tokyo, Kashiwa, Chiba 277-8581, Japan}

\author{Ryotaro~Arita}
\affiliation{RIKEN Center for Emergent Matter Science (CEMS), Wako, Saitama 351-0198, Japan}
\affiliation{Department of Applied Physics and Quantum-Phase Electronics Center (QPEC), The University of Tokyo, Tokyo 113-8656, Japan}

\author{Keji~Lai}
\affiliation{Department of Physics, The University of Texas at Austin, Austin, Texas 78712, United States}

\author{Takao~Sasagawa\text{*}}
\affiliation{Materials and Structures Laboratory, Tokyo Institute of Technology, Yokohama, Kanagawa 226-8503, Japan}

\author{Takeshi~Kondo\text{*}}
\affiliation{Institute for Solid State Physics, University of Tokyo, Kashiwa, Chiba 277-8581, Japan}

\date{\today}

\maketitle

{\bf The van der Waals (vdW) materials with low dimensions have been extensively studied as a platform to generate exotic quantum properties \cite{Cao2018a,Cao2018b,Tran2019,Alexeev2019,Seyler2019,Jin2019a}. Advancing this view, a great deal of attention is currently paid to topological quantum materials with vdW structures.
Here, we provide a new concept of designing topological materials by the vdW stacking of quantum spin Hall insulators (QSHIs). Most interestingly, a slight shift of inversion center in the unit cell caused by a modification of stacking is found to induce the topological variation from a trivial insulator to a higher-order topological insulator (HOTI).
Based on that, we present the first experimental realization of a HOTI by investigating a bismuth bromide Bi$_4$Br$_4$ \cite{VonBenda1978,Filatova2007,Zhou2014a,Li2019e,Tang2018e,Hsu2019b} with angle-resolved photoemission spectroscopy (ARPES). The unique feature in bismuth halides capable of selecting various topology only by differently stacking chains, combined with the great advantage of the vdW structure, offers a fascinating playground for engineering topologically non-trivial edge-states toward future spintronics applications.}

The $Z_2$ weak topological insulator (WTI) phases have been confirmed in the materials with stacked QSHI layers, where the side-surface becomes topologically non-trivial by accumulating helical edge states of QSHI layers \cite{Rasche2013,Noguchi2019}. Similarly, higher-order topological insulators (HOTIs) are expected to be built from stacking QSHIs, which, however, accumulate the 1D edge-states to develop 1D helical hinge-states in a 3D crystal \cite{Khalaf2018,Matsugatani2018a}. Such HOTI phases have been theoretically predicted recently in materials previously regarded as trivial insulators under the $Z_2$ criterion by extending the topological classification to the $Z_4$ topological index \cite{Po2017,Song2018b,Zhang2018k,Vergniorya,Tang2019a,Khalaf2018PRX}. To date, only one material has been experimentally confirmed to be in the higher-order topological phase, which is bulk bismuth \cite{Schindler2018}. However, bulk bismuth is a semimetal, which cannot become insulating even by carrier doping. Materials science is, therefore, awaiting the first experimental realization of a HOTI, which enables one to explore various quantum phenomena including spin currents around hinges and quantized conductance under the external fields.

A quasi-1D bismuth bromide, Bi$_4$Br$_4$, with a bilayer structure of chains (Fig. 1b) is theoretically predicted to be a topological crystalline insulator of $Z_{2,2,2,4}=\{0,~0,~0,~2\}$, protected by the $C_2$-rotation symmetry  \cite{Zhang2018k,Vergniorya,Tang2019a,Tang2018e,Hsu2019b,Li2019e}. This state should develop 2D topological surface states in the cross-section (010) of the chains  \cite{Fang2019,Zhang2019d}. Significantly, theory also categorizes this system as a HOTI, and expects that 1D helical hinge-states emerge between the top-surface (001) and the side-surface (100) of a crystal due to the second-order bulk-boundary correspondence \cite{Tang2018e,Hsu2019b}. Nevertheless, the topological phase of Bi$_4$Br$_4$ has not been experimentally revealed so far, and thus spectroscopic measurements are necessary to identify that Bi$_4$Br$_4$ is indeed a HOTI.

Here we propose that the quasi-1D bismuth halides \cite{VonBenda1978,Filatova2007} offers an
excellent platform to realize various topological phases selected by different stacking structures of the Bi$_4$$X_4$ ($X$=I or Br) chains (Figs. 1a-d). A WTI state emerges in $\beta$-Bi$_4$I$_4$ with single-layered chains per unit cell (A-stacking; Fig. 1c), in which quasi-1D topological surface states have been observed in the side surface of the crystal \cite{Noguchi2019}. A trivial insulator phase is instead obtained in $\alpha$-Bi$_4$I$_4$, where the chains take a double-layered structure  (AA'-stacking; Fig. 1d). While a HOTI candidate Bi$_4$Br$_4$ also consists of double layers, one of these is flipped by 180$^\circ$ in the unit cell (AB-stacking; Fig. 1b), indicating that the difference in stacking induces the topological phase variation from a trivial insulator to a HOTI.

To understand the difference of the topological class between the AA'- and AB-stacking, we have conducted DFT calculations for $\alpha$-Bi$_4$I$_4$ and Bi$_4$Br$_4$ (Figs. 1g and h, respectively).
A critical difference between the two compounds is seen in the order of parity eigenvalues before the inclusion of the spin-orbit coupling (SOC) at the L point (left panels in Figs. 1g and h): the bilayer-split states near $E_{\rm F}$ of Bi$_4$Br$_4$ have the same signs in the parities, $\rm{(--)}$ [$\rm{(++)}$], for the bulk valence bands (BVBs) [the bulk conduction bands (BCBs)]. This contrasts to $\alpha$-Bi$_4$I$_4$ with the opposite signs, $\rm{(+-)}$  [$\rm{(-+)}$], for BVBs [BCBs]. The inclusion of SOC induces a band inversion between BVBs and BCBs (right panels in Figs. 1g and 1h). In $\alpha$-Bi$_4$I$_4$, the parities of BVBs stay opposite without changing the overall parity of the occupied states; this compound is, therefore, categorized as a trivial insulator. In contrast, the parities of BVBs are varied from $\rm{(--)}$ to $\rm{(++)}$ in Bi$_4$Br$_4$ due to a double band inversion (Fig. 1f) between the pair of BVBs and that of BCBs; this leads to a HOTI state with $Z_4=2$, which is never realized in $\alpha$-Bi$_4$I$_4$. Another phase of $Z_2$ topological insulator, instead of a HOTI, could be realized in Bi$_4$Br$_4$ when SOC is not strong enough to fully induce the double band inversions, mixing BVBs and BCBs only subtly with each other. This possibility can be, however, easily denied experimentally by observing the band gap ($E_{\rm gap}$) which is much larger than the bilayer splitting ($E_{\rm split}$), as indeed we have confirmed (and will demonstrate later) in our ARPES measurements. In passing, the band inversion at M is not critical for the bulk topology in both the compounds when SOC is large enough, since it does not affect the overall parity of the occupied bands.

For further comparison, we have also calculated the Bloch functions (Supplementary Note 1) and found that a slight shift of the inversion center in the unit cell, when changing from the AA'- to AB-stacking, achieves the  order of parity eigenvalues required for a HOTI. In Figs. 1i and j, we illustrate the parity eigenvalues of
BVBs and BCBs at the L point, which are both split to the bonding and antibonding states due to bilayer structure. The inversion center in the AA'-stacking ($\alpha$-Bi$_4$I$_4$ case) is located between two layers (a star in Fig. 1d and Fig. 1i). In this circumstance, the wave functions have opposite parities both for BVBs and BCBs, $\rm{(+-)}$ and $\rm{(-+)}$, respectively. In the AB-stacking (Bi$_4$Br$_4$ case), the inversion center is shifted to inside of a plane (a star in Fig. 1b and Fig. 1j). Significantly, this shift changes the parities of bilayer-split bands to the same sign, $\rm{(--)}$ and $\rm{(++)}$, for BVBs and BCBs, respectively, which yields a HOTI phase under a strong SOC. The variation of parities caused by the different stacking (AA’- and AB-stacking) is illustrated also in Supplementary Fig. S2.

We have analyzed the stacking-dependent topological properties by nanowire calculations (Figs. 1k and l; see Method), which will also give a clue for the future engineering of topological vdW chains. The gap opens at $E_{\rm F}$ in the $\alpha$-Bi$_4$I$_4$ nanowire (Fig. 1k), being consistent with the trivial insulating phase. In contrast, gapless topological hinge states emerge in the Bi$_4$Br$_4$ nanowires, where the valence and conduction bands are connected by a Dirac-like dispersion (Fig. 1l). These gapless states emerge due to the nontrivial higher-order topology in Bi$_4$Br$_4$ and are attributed to the topological hinge states. Since the bulk crystal has the same symmetry as the nanowire we set for calculations, the protected hinge states should also emerge in a 3D crystal.

Figure 2a shows a photographic image of Bi$_4$Br$_4$ crystals we used for experiments.
XRD measurements have confirmed a bilayer structure in the crystal (Supplementary Note 3).
A semiconducting property is observed in the overall behavior of electrical resistivity (Fig. 2b).
Significantly, however, we find a saturation toward the lowest temperature, implying that topological edge-states contribute to the electrical conduction. Cleaved surfaces were investigated by scanning electron microscopy (SEM) and laser microscopy (Figs. 2c and 2d, respectively). A number of terraces and facets were observed to be aligned parallel to the chain direction (the $b$-axis in Fig. 1b).
These have been revealed to be composed of the (001) and (100) crystal planes by grazing-incidence small-angle X-ray-scattering (GISAXS) measurements (Fig. 2e and Supplementary Note 4), which exhibits two reflection lines separated by $\sim 73^{\circ}$, same as the edge angle in the unit cell (see Fig. 1b).  A huge number of crystal hinges, therefore, should be naturally exposed on a cleaved surface (Fig. 2f). The local conductivity maps obtained by microwave impedance microscopy (MIM) \cite{Lai2008} also show signatures of conductive hinge states in exfoliated flakes of Bi$_4$Br$_4$ (Supplementary Note 5). This unique situation gives us a great opportunity to directly observe the topological hinge states by ARPES with a  photon beam of $\sim$50 $\mu$m in size.

We have performed synchrotron-based ARPES measurements on the cleaved surface of Bi$_4$Br$_4$ and observed a semiconducting band structure consistent with DFT band calculations (Supplementary Note 6). Figures 3a-c plot the ARPES intensities at different binding energies measured along a $k_y$-$k_z$ sheet around the $\bar{\Gamma}$ point;  $k_y$ corresponds to the chain direction, and  $k_z$ were swept by changing photon energies ($h\nu$s). While there are no states observed at $E_{\rm F}$ (Fig. 3a), line-like structures appear at the energies lower than $E-E_{\rm F}=-0.6$ eV (Figs. 3b and c). The 1D dispersions indicate that the interlayer coupling of chains along the $c$-axis is negligible (Supplementary Note 7). The electronic states along a $k_y$-$k_x$ sheet have also been measured at $h\nu=100$ eV (Fig. 3d-f). While the photoelectron intensities are weak, island-like intensities are seen around $\bar{\rm M}$ at $E_{\rm F}$ (Fig. 3d).  At higher binding energies (Figs. 3e and 3f), highly anisotropic structures appear, indicating that the band disperses weakly in the direction perpendicular to the chain (or $k_x$ direction).
These features are more directly revealed in Figs. 3g-i, by plotting the ARPES dispersion along $\bar{\rm M}$-$\bar{\rm \Gamma}$-$\bar{\rm M}$ (Fig. 3g) and along $k_y$ across $\bar{\rm \Gamma}$ (Fig. 3h) and $\bar{\rm M}$ (Fig. 3i). The valence band maximum is located about $E-E_{\rm F}=-0.3$ eV as examined in Fig. 3j, where energy distribution curves (EDCs) close to $\bar{\rm M}$ are extracted from Fig. 3i.
The weak intensities near $E_{\rm F}$, yielding islands around $\bar{\rm M}$ in Fig. 3d,
mostly comes from spectral tails of the unoccupied conduction band in bulk, which may be shifted down to lower binding energies by the effect of surface band bending. While the overall band shape is similar to that of  $\alpha$-Bi$_4$I$_4$ \cite{Noguchi2019}, the semiconducting gap in Bi$_4$Br$_4$ ($\sim 0.3$ eV) is more than two times larger (Supplementary Note 8). These results agree with our DFT calculations, which expect that the band gap calculated with SOC is larger in Bi$_4$Br$_4$ than $\alpha$-Bi$_4$I$_4$ (see Figs. 1g and h).
Since the band structure is very simple and the band gap is relatively large for topological insulators, Bi$_4$Br$_4$ is an ideal candidate for the first demonstration of a HOTI state.

Here we provide evidence for the topological hinge states in Bi$_4$Br$_4$ obtained by a laser-based ARPES that has many advantages to investigate topological states, owing to its high energy- and momentum-resolution and high efficiency of data acquisition \cite{Lv2019a}.
Figure 4a plots the ARPES intensities at $E_{\rm F}$ covering a rectangle region in the upper panel; relatively weak signals are clearly seen by selecting a proper color scale.
Island-like intensities around $\bar{\rm M}$ originate from the spectral tails, as argued above. More importantly, our high-quality data clearly exhibit intensities with a 1D distribution along the line at $k_y=0$.
In Fig. 4d, we plot ARPES dispersions obtained along several momentum cuts (orange dashed lines in Fig. 4a). The hole-like bands approach $E_{\rm F}$ at $\bar{\rm M}$, and disperse toward higher binding energies as moving away from $\bar{\rm M}$. Detailed properties around $\bar{\rm M}$ are examined in Fig. 4e by extracting EDCs; these have a good agreement with those of the bulk bands observed by synchrotron-based ARPES, showing the band gap ($E_{\rm gap}$) of $\sim 0.3$ eV and finite intensities at $E_{\rm F}$ due to the spectral tails of the unoccupied conduction band. We have also more directly estimated the value of $E_{\rm gap}\sim0.3$ eV by the experiment with the K-deposition technique on the sample surface, which brings the conduction bands down below $E_{\rm F}$ (see supplementary Fig. S8).

Most interestingly, we found metallic in-gap states with Dirac-like dispersions for all the momentum cuts, as clearly visible by changing the color scale (Fig. 4c); the 1D Fermi surface in Fig. 4a is, thus, formed by these dispersions. The in-gap states are also confirmed in the momentum distribution curves (MDCs) at the roughly estimated Dirac point ($E-E_{\rm F} = -60$ meV)  (Fig. 4b). Magnified ARPES images (marked by yellow boxes in Figs. 4c and 4d) with higher statistics (Fig. 4f) and a curvature plot for one of these (Fig. 4g) \cite{Zhang2011} visualize the Dirac band inside the bulk band gap. These observations are in sharp contrast to the case for a trivial insulator $\alpha$-Bi$_4$I$_4$ (Figs. 4h-j): there is no indication for in-gap states in the image (Fig. 4j) even after changing the color scale (Fig. 4i), nor in MCD at $E-E_{\rm F}=-60$ meV (Fig. 4h). The Dirac band we observed in Bi$_4$Br$_4$, therefore, is attributed to the topological state. This result is especially intriguing in that the difference in chain-stacking is the origin of the different band topologies between Bi$_4$Br$_4$ and $\alpha$-Bi$_4$I$_4$. The experimentally determined $E_{\rm gap}$ ($\sim0.3$ eV) of Bi$_4$Br$_4$ is much larger than the bilayer splitting $E_{\rm split}$ ($\sim0.1$ eV), avoiding a rare situation that BVBs and BCBs are mixed only subtly. 
Therefore, the 1D Dirac-band we observed should be due to a topological hinge state, and thus Bi$_4$Br$_4$ is a HOTI.

The noble concept of the topological material-design we have provided will open up a new pathway to materials science.
In particular, the Bi$_4$Br$_4$ verified to be the first higher-order topological insulator provides various advantages in application, owing to its unique crystal structure built from the van der Waals stacking of bismuth bromide chains; since a bunch of crystal hinges are exposed on this crystal, a large amount of topological edge-current could be generated.  In bismuth halides, multiple topological phases can be selected by different procedures of chain-stacking, which offers the capability of designing and engineering novel functional materials by thin film growth and microfabrication. Notably, the 1D nanowires obtained by simple exfoliation technique may play a crucial role in developing the devices for Majorana-based topological computing by transferring the nanowires onto a superconducting substrate \cite{Hsu2018a,Jack2019}.

{\bf  Data availability}\\
The data that support the findings of this study are available from the corresponding author on reasonable request.

{\bf  Acknowledgements}\\
 We thank D. Hamane for SEM characterization of the sample surface. We also thank X. Ma and D. Abeysinghe for their support in the exfoliation of Bi$_4$Br$_4$ samples. The work done at Tokyo Institute of Technology was supported by a CREST project [JPMJCR16F2] from Japan Science and Technology Agency (JST). The GISAXS experiments were performed under the approval of PF-PAC No. 2018G661. We thank Diamond Light Source for access to beamline I05 under proposal SI20445 that contributed to the results presented here. Use of the Stanford Synchrotron Radiation Light source, SLAC National Accelerator Laboratory, is supported by the U.S. Department of Energy, Office of Science, Office of Basic Energy Sciences under Contract No. DE-AC02-76SF00515. The MIM work was supported by the United States Army Research Office under Grant No. W911NF-17-1-0542. R.N. acknowledges support by JSPS under KAKENHI Grant No.18J21892 and support by JSPS through the Program for Leading Graduate Schools (ALPS).

{\bf  Author Contribution}\\
	T.Ko. and T.Sa. planned the experimental project.
	R.N. conducted ARPES experiments and analyzed the data.
	Ke.K, P.Z., C.L., C.B., S.Sa., H.T., S.K., Ki.K.,K.Y., A.H., V.K., A.G., A.B., T.Ki., C.C., M.H., D.L., S.Sh., and T.Ko. supported ARPES experiment.
	R.N., Z.J., Z.X., D.L. and K.L. performed MIM experiments and analyzed the data.
	M.K., T.T. and T.Sa. made and characterized Bi$_4$Br$_4$ single crystals and
	 performed transport experiments. R.N. and M.K. took laser-microscope images. T.Sh. performed GISAXS experiment. R.N. measured SEM image.
	T.Sa., M.H., M.O. and R.A. calculated the band structure and analyzed the band topology.
	R.N., Z.J., Ke.K., M.H., M.O., T.Sh., K.L., T.Sa. and T.K. wrote the paper.
	All authors discussed the results and commented on the manuscript.

{\bf  Author Information}\\
The authors declare no competing financial interests. Correspondence and requests for materials should be addressed to T.Sa~(email: sasagawa@msl.titech.ac.jp) or T.K.~(email: kondo1215@issp.u-tokyo.ac.jp).

\newpage
{\bf METHODS}

{\bf Samples.}\\
Following a similar procedure as described in \cite{Noguchi2019}, single crystals of Bi$_4$Br$_4$ were grown by the chemical vapor transport method. The transport agent used was HgBr, and the optimized temperatures were 285 $^\circ$C and 188 $^\circ$C for the source and growth zones, respectively. The electrical resistivity of Bi$_4$Br$_4$ single crystals was measured by a standard four-probe technique using the Quantum Design Physical Property Measurement System (PPMS).

{\bf  DFT calculations.}\\
First, we performed the structural optimization using the Perdew-Burke-Ernzerhof (PBE) parametrization of the generalized gradient approximation (GGA)~\cite{pbegga}, and the projector augmented wave method~\cite{paw} with the inclusion of the spin-orbit coupling as implemented in {\it Vienna ab initio Simulation Package}~\cite{vasp1,vasp2,vasp3,vasp4}.
In the structural optimization, we optimized atomic coordinates using the experimental lattice parameters as reported in \cite{Noguchi2019} for $\alpha$-Bi$_4$I$_4$ and shown in Fig.~S3 for Bi$_4$Br$_4$.

Next, we performed first-principles band-structure calculation using the Becke-Johnson potential~\cite{BJ} as implemented in the \textsc{WIEN2k} code~\cite{wien2k}.
The muffin-tin radii $r$ for all atoms were set to 2.5 a.u.~and the maximum modulus for the reciprocal lattice vectors $K_{\mathrm{max}}$ was chosen so that $rK_{\mathrm{max}}=8$.
We took a 18$\times$18$\times$3 $\bm{k}$-mesh.
From the calculated band structures, we extracted the Wannier functions of the Bi-$p$ and Br(I)-$p$ orbitals using the \textsc{Wien2Wannier} and \textsc{Wannier90} codes~\cite{Wannier1,Wannier2,Wien2Wannier,Wannier90}.
We did not perform the maximal localization procedure for the Wannier functions to prevent orbital mixing among the different spin components.
In the nanowire calculations for Bi$_4$Br$_4$ and $\alpha$-Bi$_4$I$_4$, we set 12$\times$5 and 12$\times$6 wires, respectively, which have the same symmetry as each bulk crystal, and used the Wannier function, where we stacked the 12 wires along the $a$-axis and 5 and 6 wires along the $c$-axis. In the 12$\times$5 Bi$_4$Br$_4$ wires, the two surfaces perpendicular to the $c$-axis are the A-layers in Fig.1.

{\bf  GISAXS measurements.}\\
GISAXS experiments were performed at the beamline 3A of the Photon Factory at KEK. The x-ray energy was 11 keV, and the slit-cut beam size was 0.1 $\times$ 0.1 mm. The glancing angle of the x-ray was 0.1$^\circ$ with respect to the Bi$_4$Br$_4$ (001) plane. The scattered x-rays were detected with PILATUS-100K located 1 m downstream of the sample. For the alignment of sample orientation, the 001 and 200 Bragg reflections were used. All the measurements were performed at room temperature.

{\bf ARPES set-up.}\\
Standard-ARPES measurement with synchrotron radiation was performed at the HR-ARPES branch of the beamline I05 of the Diamond Light Source equipped with a ScientaOmicron R4000 analyzer and beamline 5-2 of Stanford Synchrotron Radiation Lightsource (SSRL) equipped with a ScientaOmicron DA30L analyzer. The  sample temperature was fixed at ~10 K during the measurements.  The photon energy was set at between 70 and 100 eV. The angular  resolution was 0.1$^{\circ}$ and the overall energy resolution was better than 20~meV.
Laser-based ARPES measurements were performed at the Institute for Solid State Physics, The University of Tokyo~\cite{yaji}.
The laser system provides 6.994-eV photons~\cite{shimojima}.
The measurement temperature of the samples was about 20~K. The angle resolution was 0.3$^{\circ}$ and the overall energy resolution was set to less than~5~meV.

Synchrotron-based nano-ARPES measurements presented in Supplementary Note 7 were performed at the 3.2L-Spectromicroscopy beamline of the Elettra Light Source~\cite{Dudin}. A Schwarzschild objective was used to focus the photon beam to a spot of less than 1 $\mu$m in size. The photoelectron detector is facility-made, and it is rotatable inside the measurement chamber, which also enables Fermi surface mapping without sample rotation. The photon energy was set at 74 eV, and the sample temperature was kept at around 20 K. The overall energy resolutions were set to be better than 60 meV.
ARPES data were analyzed using the inverse mapping functions described in \cite{Ishida}.

{\bf MIM measurements and sample preparations.}
The MIM work presented in Supplementary Note 5 was performed on a commercial AFM platform (ParkAFM XE-70). Shielded cantilever probes from PrimeNano Inc. were used. Bi$_4$Br$_4$ thin flakes used in MIM experiments were first mechanically exfoliated onto polypropylene carbonate (PPC)-coated polydimethylsiloxane (PDMS) stamps using Scotch tape. The flakes were then transferred onto SiO$_2$ (300 nm)/Si substrates and immediately moved to the measurement systems.



\clearpage
\begin{figure*}
\includegraphics[width=5.3 in]{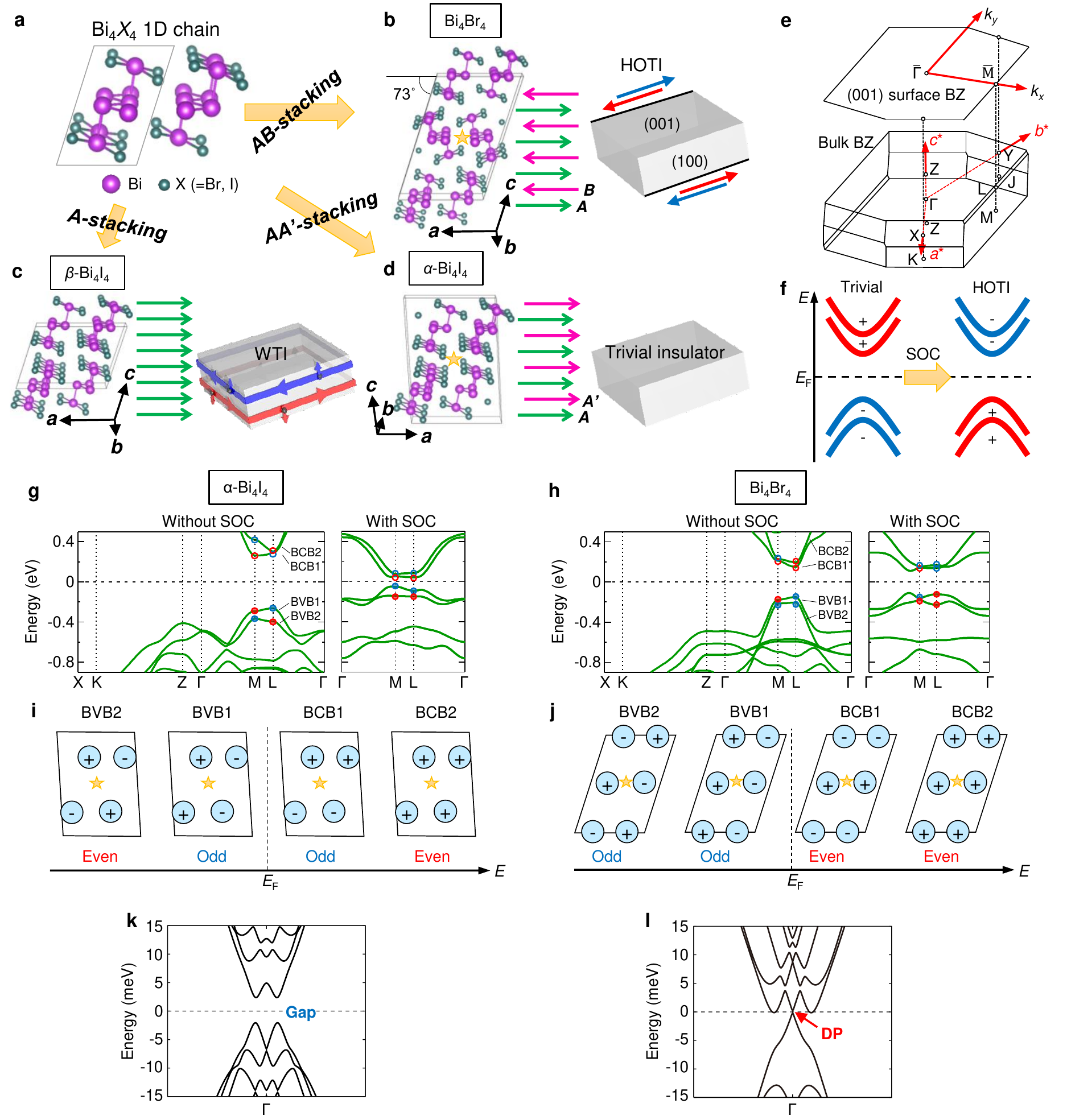}
\renewcommand{\baselinestretch}{1.2}
\caption{\textbf{Stacking-dependent topological properties in Bi$_4X_4$ ($X$=Br, I).}
\textbf{a,} Crystal structure of Bi$_4X_4$ 1D chain. \textbf{b-d,} Crystal structures and schematics of topological phases of Bi$_4$Br$_4$ ({\bf b}), $\beta$-Bi$_4$I$_4$ ({\bf c}) and $\alpha$-Bi$_4$I$_4$ ({\bf d}) viewed from the chain direction. The magenta and green arrows indicate the stacking sequence of Bi$_4X_4$ chains. The stars in {\bf b} and {\bf d} indicate the positions of the inversion center. \textbf{e,} The Brillouin zone (BZ) for the bulk and the projected surface of Bi$_4$Br$_4$. \textbf{f,} Schematic of a double band inversion that induces a higher-order topological insulator (HOTI) phase. \textbf{g, h,}  Bulk band structures calculated for $\alpha$-Bi$_4$I$_4$ ({\bf g}) and Bi$_4$Br$_4$ ({\bf h}) without the spin-orbit coupling (SOC) and with SOC around the Fermi energy ($E_{\rm F}$). The red and blue circles label the even parities and the odd parities, respectively. \textbf{i, j,} Schematics of the wave functions and the corresponding parity eigenvalues (denoted as even or odd) for the bulk valence bands (BVBs) and the bulk conduction bands (BCBs) close to $E_{\rm F}$ at the L point of $\alpha$-Bi$_4$I$_4$ ({\bf i}) and Bi$_4$Br$_4$ ({\bf j}) before the inclusion of SOC.
\textbf{k, l,} Band calculations of $\alpha$-Bi$_4$I$_4$ nanowires (12$\times$6 wires) and Bi$_4$Br$_4$ nanowires (12$\times$5 wires) with the same symmetry as each balk crystal, respectively.
 Band gap is opened for $\alpha$-Bi$_4$I$_4$, whereas gapless topological hinge states emerge for  Bi$_4$Br$_4$; the Dirac point is indicated by the red arrow in {\bf l}. The unit cell used for counting the number of the Bi$_4X_4$ wires is indicated by the gray parallelogram in {\bf a}.
}
\label{fig1}
\end{figure*}

\clearpage
\begin{figure*}
\includegraphics[width=6.2in]{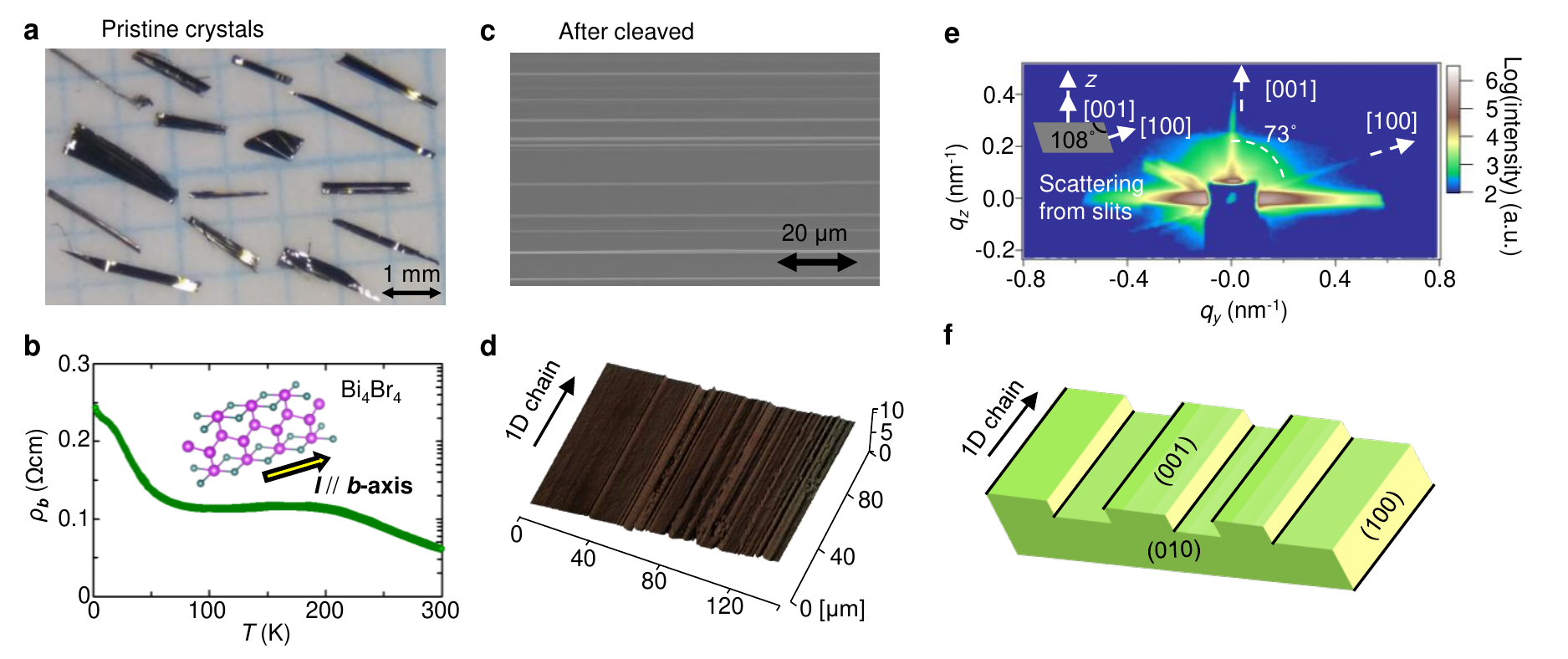}
\renewcommand{\baselinestretch}{1.2}
\caption{\textbf{Hinges in the cleaved surface of semiconducting Bi$_4$Br$_4$.}
\textbf{a,} Photographic image of Bi$_4$Br$_4$ crystals. \textbf{b,} Temperature dependence of the electrical resistivity measured along the chain direction ($\rho_b$). \textbf{c,} Scanning electron microscope (SEM). \textbf{d,} Laser-microscope images of cleaved surfaces. \textbf{e,} Results of grazing-incidence small-angle X-ray-scattering (GISAXS) measurement of a cleaved Bi$_4$Br$_4$. \textbf{f,} Schematic of a cleaved surface with the (001) plane and the (100) plane. The black lines show the hinges between these two planes,  where the topological hinge states are expected to emerge.
}
\label{fig2}
\end{figure*}
\clearpage

\begin{figure*}
\includegraphics[width=6.5in]{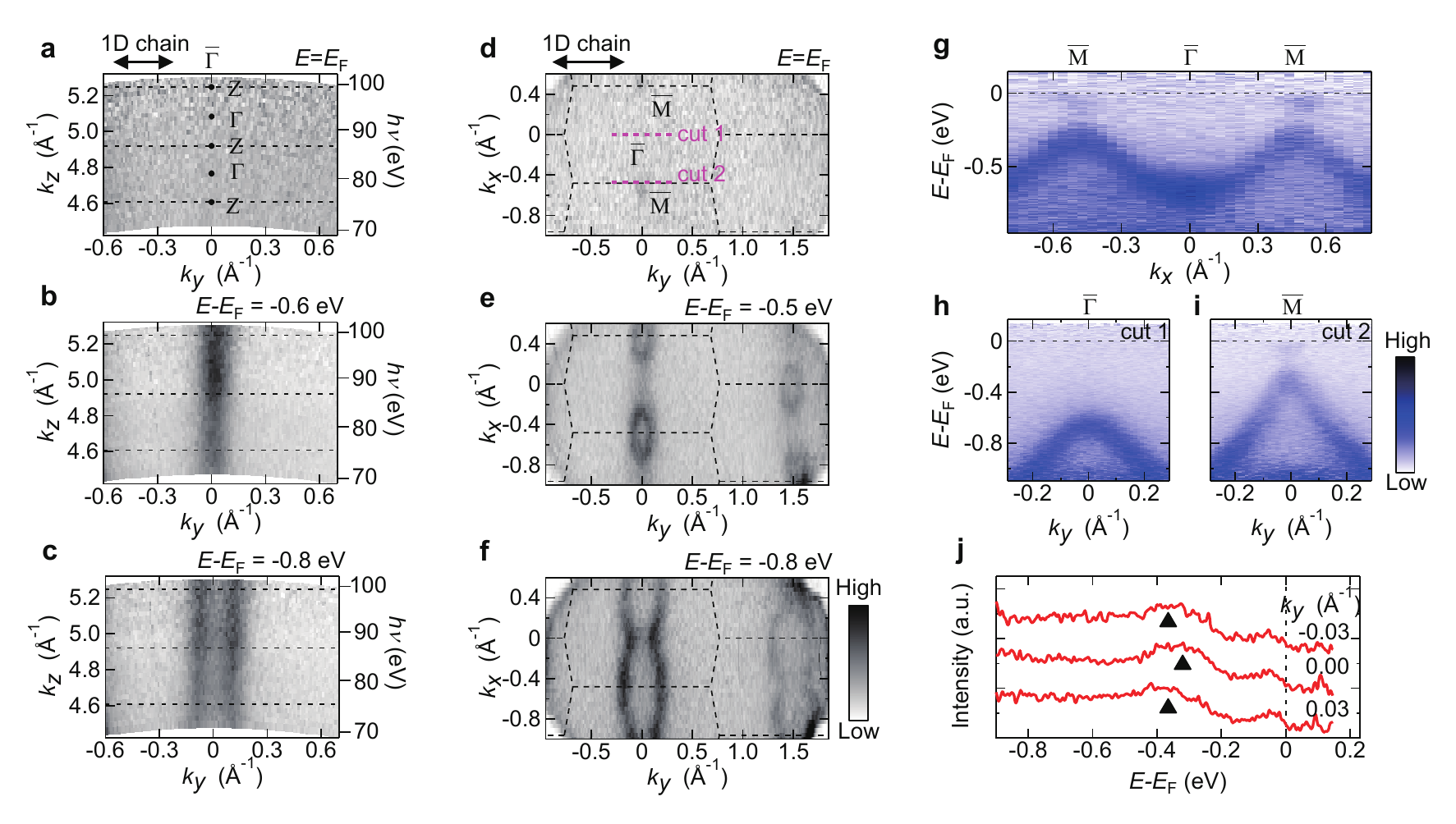}
\renewcommand{\baselinestretch}{1.2}
\caption{\textbf{ARPES mapping for bulk band structure of Bi$_4$Br$_4$}  \textbf{a-c,} ARPES map along
a $k_y$-$k_z$ sheet plotted at different energies of  $E=E_{\rm F}$ ({\bf a}), $E-E_{\rm F}=-0.6$ eV ({\bf b}) and $E-E_{\rm F}=-0.8$ eV ({\bf c}). $k_y$ and $k_z$ correspond to the directions parallel and perpendicular to the 1D chain, respectively.
The data for the different $k_z$ values were measured by changing $h\nu$ from 70 to 100 eV. The black dashed lines denote constant $k_z$, cutting across the Z points of the Brillouin zone. The ARPES intensities are integrated within a $\pm$25~meV window about each binding energy. \textbf{d-f,} ARPES map along a $k_y$-$k_x$ sheet plotted at different energies of  $E=E_{\rm F}$ ({\bf d}), $E-E_{\rm F}=-0.5$ eV ({\bf e}) and $E-E_{\rm F}=-0.8$ eV ({\bf f}). $k_y$ and $k_x$ correspond to the directions parallel and perpendicular to the 1D chain, respectively.  The data were measured at $h\nu =100$ eV. The black dashed lines indicate the surface Brillouin zones of the (001) surface. The intensities are integrated within a $\pm$25~meV window about each binding energy. \textbf{g,} ARPES band dispersion along the $\bar{\rm M}$-$\bar{\rm \Gamma}$-$\bar{\rm M}$ cut, which is perpendicular to the 1D chain direction. \textbf{h, i,} ARPES band dispersions in the 1D chain direction, measured along cut 1 and cut 2 in {\bf d} (magenta dashed lines). \textbf{j,} Energy distribution curves around the $\bar{\rm M}$ point extracted from {\bf i}. The black markers indicate the energy positions for the valence band closest to $E_{\rm F}$.
}
\label{fig3}
\end{figure*}

\clearpage
\begin{figure*}
\includegraphics[width=6.5in]{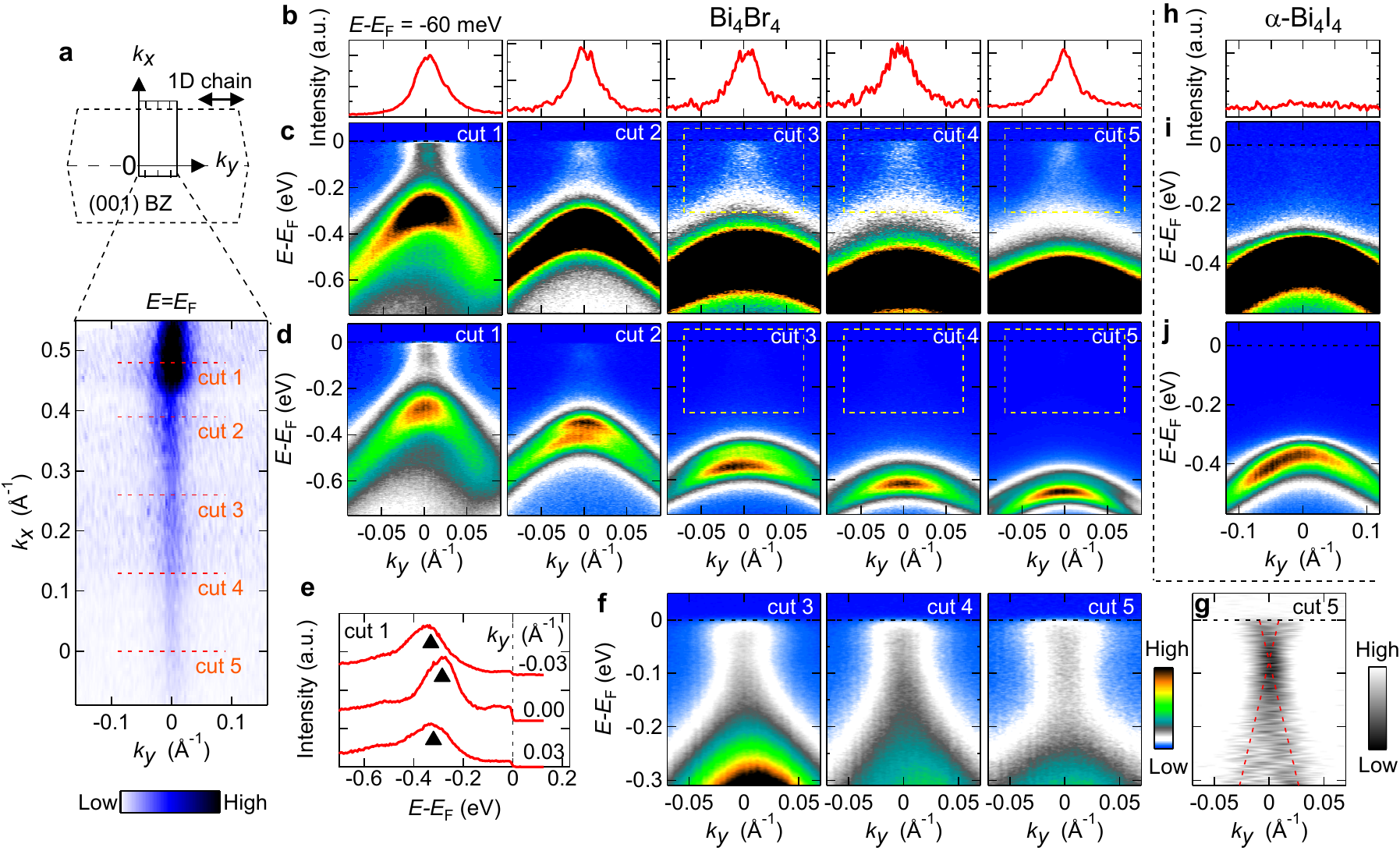}
\renewcommand{\baselinestretch}{1.2}
\caption{\textbf{Experimental band structures of topological hinge states by laser-ARPES.}
\textbf{a,} ARPES intensity distributions in the $k_x-k_y$ plane at $E_{\rm{F}}$ of Bi$_4$Br$_4$. $k_y$ and $k_x$ correspond to the directions parallel and perpendicular to the 1D chain, respectively.  The intensities are integrated within a $\pm$20~meV window about the Fermi level.
The black dashed lines indicate the surface Brillouin zone for the (001) surface.
\textbf{b-d,} APRES band maps measured at various momentum cuts indicated in {\bf a} ({\bf c}), the same maps shown with enhanced color contrasts ({\bf d}), and the corresponding momentum distribution curves at $E-E_{\rm F}=-60$ meV ({\bf b}). 
\textbf{e,} Energy distribution curves around the $\bar{\rm M}$ point extracted from the left panel (cut 1) in {\bf d}. The black markers indicate the energy positions for the valence band closest to $E_{\rm F}$.
\textbf{f,} Magnified ARPES band maps taken at cut 3-5 in {\bf d}. Energy and momentum ranges are indicated by the yellow dotted rectangles in {\bf c} and {\bf d}. 
\textbf{g,} Curvature plot of ARPES band map for cut 5 in {\bf f}. The red line is a guide to the eyes to trace the Dirac-like linear dispersion. \textbf{h-j,} ARPES band maps of $\alpha$-Bi$_4$I$_4$ around the $\bar{\Gamma}$ point ({\bf j}), the same map shown with enhanced color contrast, and the momentum distribution curve at $E-E_{\rm F}=-60$ meV ({\bf h}). The momentum region for  \textbf{h-j }
corresponds to cut 5 for the Bi$_4$Br$_4$ data shown in the right panels of {\bf c} and {\bf d}.
The data of {\bf h-j} for $\alpha$-Bi$_4$I$_4$ are reproduced from \cite{Noguchi2019} to compare with the current results of Bi$_4$Br$_4$.
}
\label{fig4}
\end{figure*}


\end{document}